# Spontaneous structural distortion of metallic Shastry-Sutherland system $DyB_4$ by quadrupole-spin-lattice coupling

Hasung Sim,[1,2] Seongsu Lee,[3] Kun-Pyo Hong,[1] Jaehong Jeong,[1,2] J. R. Zhang,[4] T. Kamiyama,[4] D. T. Adroja,[5,6] C. A. Murray,[7] S. P. Thompson,[7] F. Iga,[8] S. Ji,[9] D. Khomskii,[10] and Je-Geun Park[1,2#]

1. Center for Correlated Electron Systems, Institute for Basic Science (IBS), Seoul 08826, Korea
2. Department of Physics & Astronomy, Seoul National University, Seoul 08826, Korea
3. Neutron Science Division HANARO, Korea Atomic Energy Research Institute, Daejeon, 305-353, Korea
4. Institute of Materials Science & J-PARC Center, KEK, Tsukuba 305-0801, Japan
5. ISIS Facility, Rutherford Appleton Laboratory, Didcot OX11 0QX, UK
6. Highly Correlated Matter Research Group, Physics Department, University of Johannesburg, Auckland Park 2006, South Africa
7. Diamond Light Source, Harwell Science and Innovation Campus, Didcot, OX11 0DE, UK
8. Department of Physics, Ibaraki University, Mito 310-0056, Japan
9. Department of Physics, Pohang University of Science and Technology, Pohang, 790-784, Korea
10. II. Physikalisches Institut, University of Koeln, 50937 Koeln, Germany

Abstract
$DyB_4$ has a two-dimensional Shastry-Sutherland (Sh-S) lattice with strong Ising character of the Dy ions. Despite the intrinsic frustrations, surprisingly, it undergoes two successive transitions: a magnetic ordering at $T_N$ = 20K, and a quadrupole ordering at $T_Q$=12.5 K. From high-resolution neutron and synchrotron X-ray powder diffraction studies, we have obtained full structural information on this material in all phases, and demonstrate that structural modifications occurring at quadrupolar transition lead to the lifting of frustrations inherent in the Sh-S model. Our study thus provides a complete experimental picture of how the intrinsic frustration of the Sh-S lattice can be lifted by the coupling to quadrupole moments. We show that two other factors, i.e. strong spin-orbit coupling and long-range RKKY interaction in metallic $DyB_4$, play an important role in this behavior.

## 1. Introduction

In a system where a strong coupling prevails among the four fundamental degrees of freedom of a solid: spin, lattice, orbital (quadrupole) and charge, our understanding based on weak coupling theories seems to break down and subsequently new physics emerges. Of recent discoveries in the field, it is particularly interesting that some strongly correlated electron systems show features that can only be understood in terms of a strong coupling between the four degrees of freedom of a solid. One of the prime interests is how this cross coupling modifies the ground states that are often of a complex nature due to competing factors.

Among the four degrees of freedom, orbital is relatively less well understood. Although it was proposed under a different name, quadrupole, to explain some of the exotic magnetic properties of rare-earth magnetic systems some thirty years ago [1], the latest revival of interests in the orbital degree of freedom owes much to the recent activities on transition metal oxides, in particular manganites [2, 3]. Perhaps one important difference between the transition metal oxides and the rare-earth systems is that because d electrons are involved more directly in bonding than f electrons, the effect of the orbital ordering is relatively more striking for transition metal oxides than for rare-earth systems. However, that does not necessarily mean that the orbital or quadrupole interaction is weak for rare-earth systems. On the contrary, a recent discovery of a rare-earth system with a strong quadrupole coupling seems to reveal new physics that has not been previously found in transition metal oxides with orbital interaction [4, 5]. Perhaps because of the less direct bonding nature of f electrons one would expect more interesting and diverse lattice models realized with quadrupole orderings, which would then challenge and so enrich our understanding of materials.

The Shastry-Sutherland (Sh-S) model is a square lattice with competing nearest-diagonal and next nearest interactions, which results in magnetic frustration when they are both of an antiferromagnetic nature [6]. As it is one of a very few exactly solvable models, it has attracted significant attention over the past decades or so with a few experimental reports of several magnetic systems described by this model [7, 8]. With the experimental demonstration of a simple Sh-S lattice fully confirmed, attention has been shifting more recently toward other questions such as how the ground state of a Sh-S lattice can be modified by a further coupling to other degrees of freedom like quadrupole moments. Another interesting, yet completely open question is how the ground state of a Sh-S lattice varies under a long-ranged interaction of itinerant RKKY type. With several new Sh-S lattice systems being found in metallic systems, we feel that this is a legitimate and equally urgent question. Rare-earth boride compounds, $RB_4$, are an ideal system to address these questions as it is a metallic Sh-S system [9, 10] with the additional possibilities of multipolar physics, which arise from the rare-earth's unquenched orbital moments [1, 11].

Early studies of rare-earth tetraborides $RB_4$ go back to the 1970s when physical properties of several them were first studied [12-15]. Subsequent neutron diffraction studies found that these $RB_4$ order magnetically at low temperatures [14]. With the revival of interest in the quadrupole ordering, $RB_4$ has attracted more attention over the past few years. In particular, $DyB_4$ was recently examined by both bulk measurements [16] and resonant X-ray scattering [17, 18], and found to have a strong quadrupole ordering. First, it has an antiferromagnetic ordering at $T_N$=20 K, and then there is a second transition at a lower temperature, $T_Q$=12.5 K. At this second transition, the X-ray scattering experiment found that it has a quadrupole ordering and a structural transition from tetragonal to possibly monoclinic structures, indicative of a quadruple-lattice coupling. This quadrupole ordering at $T_Q$ is mainly due to the ordering of the quadrupole moment (asymmetric charge distribution) of the ground state of Dy.

Despite its importance related to the quadrupole physics via a quadrupole-strain coupling, one should note that the crystallographic structure, in particular in the quadrupole-ordered phase, is not sufficiently accurately known. For example, our earlier work reported in Ref. 18 was based on the collection of few Bragg peaks in single crystal X-ray scattering. So the question of the low-temperature crystal structure warrants new investigations including full structure refinements, which is the main motivation of this work.

Of further interest is that Dy with a strong Ising spin character forms the Sh-S lattice, which is known to have strong intrinsic frustration [6]. Therefore, it offers a unique opportunity, where both spin and quadrupole degrees of freedom are confined to a lattice having an intrinsically strong frustration. Because of the strong spin-orbit interaction of relativistic origin for Dy 4f electrons and the natural coupling of the quadrupole moment to the lattice, one may expect to see a spontaneous symmetry breaking of the Sh-S lattice of Dy moments right at $T_Q$, making it a rare case of such a phenomenon. A final comment to be noted is that all this interesting physics is expected to be mediated through the sea of the conduction electrons, i.e. a RKKY-type interaction.

Here, we report high-resolution neutron & synchrotron X-ray powder diffraction studies using state-of-the-art instruments where we could determine the crystallographic structure with high accuracy. In particular, we could examine how the magnetic and quadrupole ordered phases are stabilized by the structural distortion.

## 2. Experimental Details

We prepared our samples with 99.99% purity $Dy_2O_3$ and 99.52% purity $^{11}B$ isotope in stoichiometric amounts and grew single crystals using an induction furnace equipped with Xenon lamps under vacuum better than 1 mbar. Since there is the possibility of a secondary $DyB_6$ phase forming in our sample, we measured its X-ray diffraction pattern, magnetization and heat capacity after the synthesis to confirm that it shows the two previously reported phase transitions of $DyB_4$ at $T_N$=20 K and $T_Q$=12.5 K without evidence of impurities. After grinding several crystals into fine powder, we carried out high-resolution neutron diffraction experiments using the S-HRPD beamline at the J-PARC, Japan, with the resolution of $\Delta d/d \cong 3.5\times10^{-4}$ from 2 to 300 K. For comparison, we undertook high-resolution synchrotron powder diffraction experiments with $\lambda$= 0.494874 Å, using the Dynaflow cryostat on beamline I11 at Diamond Light Source [19]. We have also measured the temperature dependence of X-ray diffraction patterns using a commercial diffractometer (Empyrean, PANalytical) from 13 to ~900 K with Cu $K_{\alpha1}$ $\lambda$=1.5406 Å. All the Rietveld refinements were done using the Fullprof program [20]. We want to stress that all our high-resolution neutron and X-ray experiments are new and have not been previously reported.

For the magnetic structure studies, we reanalyzed the data that were reported in Ref. 18. These data were previously taken using a high-resolution powder diffractometer (HRPD) at the Korea Atomic Energy Research Institute (KAERI) with a Ge (331) monochromator, $\lambda$= 1.831Å [21]. Unlike our previous discussion in Ref. 18, in this study we could refine the magnetic structure using the accurate crystal structure, which was determined from the high-resolution neutron and X-ray diffraction data mentioned above. For the magnetic structure studies, we obtained all the possible magnetic structures that are allowed within a group theory using two software programs: MODY [22] and SARAh [23], and checked all the possible structures against our data.

## 3. Results and Analysis

Our S-HRPD results in the paramagnetic phase show that $DyB_4$ forms in the tetragonal structure (P 4/m b m). This tetragonal structure remains undistorted even in the magnetic phase above $T_Q$. However, once it gets cooled below $T_Q$ then there are clear signs of a structural distortion. As we present in Fig. 1, several nuclear Bragg peaks are split into new superlattice peaks. For example, the Bragg peaks of (2 1 3), (4 2 2), (3 3 2) and (4 1 2) become clearly split into two or more peaks below $T_Q$. Although our data cover a wide range of Q from 10.46 to 0.3 Å$^{-1}$, we mostly used the data points with Q larger than 3.14 Å$^{-1}$ (equivalent d-spacing is 2 Å) for our analysis in order to determine the crystal structure more accurately: $Q = \frac{4\pi \sin\theta}{\lambda} = \frac{2\pi}{d}$, where 2θ is the diffraction angle, λ the X-ray wavelength and d the spacing of (h k l) planes. Due to the Q-dependence of the magnetic form factor of the Dy 4f electrons, the magnetic peaks become considerably weaker for larger Q values so our analysis including the magnetic peaks does not improve the analysis of the crystal structure for the S-HRPD data. Thus the results reported below were obtained from the analysis done with nuclear peaks only, unless stated otherwise.

While refining our data using several possible crystal structures, we found that it can be best fitted with a monoclinic structure (P 1 21/a 1 or P 21/c in conventional notation): this low-temperature space group breaks the mirror symmetry of the paramagnetic phase. A schematic of the monoclinic structure is shown in the inset of Fig. 1(a). The summary of the structural refinement is given in Table 1, in which we list all the atomic positions as well as the lattice parameters at three representative temperatures. This result is also confirmed by our subsequent high-resolution synchrotron X-ray powder diffraction experiment results as shown in Fig. 2. In order to investigate the wider temperature dependence of the crystal structure, we carried out a further X-ray powder diffraction (XRD) experiment using a commercial x-ray diffractometer from 12 to ~900 K. As shown in Fig. 2, both X-ray and neutron diffraction data show good agreement with one another in the temperature range where they overlap.

Several things should be noted here: First, there is a clear splitting of a and b lattice constants below $T_Q$, where the monoclinic β angle deviates from 90°. Another noteworthy point is that the unit cell volume follows the usual Debye-Grüneisen formula, shown as a line in Fig. 2d, before deviating below around 150 K. In the analysis of the Debye-Grüneisen formula shown for thermal expansion, the temperature dependence of the unit cell volume is described by $V(T) = V_0\left[1 + \frac{E(T)}{Q - bE(T)}\right]$, where $V_0$ is the unit cell volume at zero temperature, $Q=(V_0B_0/\gamma)$, and $b=(B'_0-1)/2$. $B_0$ is the zero-temperature isothermal bulk modulus with $B'_0$ its first derivative with respect to pressure and γ the thermal Grüneisen parameter. The internal energy due to lattice vibrations, $E(T)$, is then given by the Debye model: $E(T) = \frac{9nk_BT}{(\theta_D/T)^3}\int_0^{\theta_D/T}\frac{x^3}{e^x-1}dx$, where $\theta_D$ is the Debye temperature, $n$ the number of atoms per unit cell, and $k_B$ the Boltzmann constant. The theoretical curve (line in Fig. 2d) was calculated by using the Debye-Grüneisen formula with the following set of parameters: $\theta_D$= 900 K, $V_0$= 203 Å$^3$, Q = 3.10×10$^{-17}$ J, and b = 1.25. This set of parameters is reasonable considering strong B-B covalent bonding when compared with other materials [24].

Subsequently, the volume continuously contracts down to 10 K, entering the quadrupole ordered phase. This indicates that there is continuous softening of the lattice down to very low temperatures, whereas normal materials usually stop thermal contraction below the boiling point of liquid $N_2$. This persistent softening of the lattice was also observed in the previous ultrasound measurements [16] and is indicative of quadrupole fluctuations existing even above $T_Q$ in $DyB_4$.

A direct consequence of the structure distortion is that there are basically four different nearest Dy-Dy distances in the quadrupole ordered phase as shown in Fig. 3a: the shortest diagonal distance (yellow), two intermediate Dy-Dy distances (red and green) and the next longer diagonal distance (blue). For comparison, the two red and green Dy-Dy distances become equal above $T_Q$. This order of Dy-Dy distance for the red and green direction is due to the fact that the angular deviation is slightly larger for the b-axis: for the red Dy-Dy distance $\theta$=149.91(7)° and for the green Dy-Dy distance $\theta'$=149.53(7)° (see Fig. 3a). As we discuss below, this splitting of Dy-Dy distances has direct implications for the magnetic ground state. In a metallic system like $DyB_4$, the magnetic exchange interaction is mediated via the RKKY interaction (J(r)) through the sea of conduction electrons: $J(r) = [\sin(2k_F r)-(2k_F r)\cos(2k_F r)]/(2k_F r)^4$. Using the Fermi wave vector appropriate for $DyB_4$, $k_F$ = 1.75 Å$^{-1}$ [14], we have plotted the Dy-Dy distance dependence of the exchange integral J(r) in Fig. 3b. Note that all Dy-Dy exchange interactions are of antiferromagnetic nature with decreasing strength in the order for the Dy-Dy distances: yellow, red, green, and blue.

Finally, we discuss the magnetic structure in the quadrupole ordered phase in more detail. A group theoretical analysis using the monoclinic structure (P 1 $2_1$/a 1) with k=(0 0 0) produces 4 different one-dimensional representations ($3\Gamma_1+3\Gamma_2+3\Gamma_3+3\Gamma_4$). Among the four representations, we found that our neutron diffraction data are more consistent with two representations ($\Gamma_2$ and $\Gamma_4$). The main difference between the two structures is along which one of the a and b axes of the monoclinic structure the antiferromagnetic interaction runs: for example, it runs along the b-axis in the $\Gamma_2$ representation while it is along the a-axis for the $\Gamma_4$ representation. If the two intermediate Dy-Dy distances are accidentally equal, then these two ground states are degenerate. However, the magnetic peaks, in particular those of (200) and (020) on one hand and (101), (0-11), (011) and (-101) on the other hand can be only properly fitted by the $\Gamma_2$ model as shown in Fig. 3(c) and (d). And this $\Gamma_2$ model is consistent with our discussion above including the RKKY interactions, i.e. *J* being larger for the red than green Dy-Dy pairs. Thus, it is indeed important to have the accurate crystal structure as we did in Fig. 1 in order to determine the magnetic structure uniquely. The temperature evolution of the magnetic Bragg peaks is shown in Fig. 3(e) and they show a clear temperature dependence across both $T_N$ and $T_Q$. The inset of Fig. 3(e) shows the temperature dependence of the components of the magnetic moments parallel to the a and c-axes: the total magnetic moment at the base temperature is 6.88(4) $\mu_B$.

## 4. Discussion and Summary

First of all, let's discuss the magnetic structure in the antiferromagnetic intermediate phase at $T_Q < T < T_N$. In this antiferromagnetic phase the in-plane components of the low-temperature magnetic phase, as shown in Fig. 3a, become zero with the moments pointing only along the c-axis. With the absence of the quadrupole ordering, the crystal structure is the same as in the paramagnetic phase (P 4/m b m). The strong Ising character of the moments is due to the crystal field splitting of Dy and the resulting strongly anisotropic g-factor [25]. In this antiferromagnetic phase, it is important to note that both red and green Dy-Dy distances are equal to one another in the tetragonal phase. This then naturally leads to strong frustration in this antiferromagnetic phase with the Sh-S lattice with Ising spins. For example, because of the strong yellow '*bond'* all Dy form antiferromagnetic pairs along the shortest bond. With the equal strength of red and green interaction, there is degeneracy in how to connect Dy-Dy along the two directions of the Dy triangle. Usually, this degeneracy produces strong frustration effects such as a large ratio of $\Theta_{CW}/T_N$, where $\Theta_{CW}$ is the Curie-Weiss temperature. Strangely enough, this ratio is unexpectedly small, ≈1.5, for $DyB_4$ with $\Theta_{CW}$ = -30 K. This may be due to additional antiferromagnetic interaction (marked by blue lines in Fig 3a), weaker than those discussed so far, which partially relieves the frustration. Even in this case, there is some residual degeneracy left in the ground state. This remaining frustration may be further released via dynamic fluctuations of the quadrupole degree of freedom, evidenced by the continuous softening of the lattice as seen in our low temperature diffraction data and the ultrasound measurements [16]. This softening of the lattice degree of freedom in the antiferromagnetic ordered phase is a direct consequence of a strong spin-orbit coupling of Dy together with quadrupole-strain coupling as discussed in Ref. 18.

However, the quadrupole ordering of Dy changes all this qualitatively. First, the magnetic moment of Dy aligned along the c-axis above $T_Q$ gets titled by 30.2(3)°. Once we have the quadrupole ordering as shown in Fig. 3a as an ellipsoid of $Q_{zx}$-type quadrupole ordering, then the directions of spins are uniquely determined by the corresponding quadrupolar orientation. That is, for Dy having a strong spin-orbit coupling and an Ising moment, the quadrupole orientation forces the spins to be parallel to the yellow bonds. This then could stabilize the magnetic structure in Fig.3a, forbidding the collinear honeycomb ordering, or even weak canting discussed above for the case of isotropic spin. It is interesting to note that the angle between the two neighboring diagonal bonds (yellow) in the quadrupole ordered phase is 90.2671°, not 90°, which is a clear, yet intriguing, indication of a subtle non-trivial distortion realized below $T_Q$.

This coupling between the spin and quadrupole degrees of freedom seems to play an important role in $DyB_4$ and, more generally, for $RB_4$. Several $RB_4$ including $TmB_4$ [10], $TbB_4$ [9] and $DyB_4$ [26] are reported to have multi-step magnetization. It is believed that these unusual metamagnetic transitions are due to the combined effects of the coupling between the two degrees of freedom too. It is even suggested that it may be related to some unusual physics found in $TmB_4$, which was ascribed to something similar to the fractional quantum Hall effect physics of a 2D electron gas [10]. Our work reported here provides the detailed structural evidence of such a coupling. Interestingly, $TbB_4$ exhibits a similar sequence of Neel and ferro-quadrupole phase transitions, with a possibly close structural distortion [27], while it appears to be absent in $TmB_4$.

We would now like to make a final comment related to oxide materials with the Sh-S lattice. For oxides, one finds the ground state of oxides more or less governed by the nearest neighbor interaction of antiferromagnetic nature. For example, an exact dimerized state is reported for $SrCu_2(BO_3)_2$ [7]. On the other hand, the physics of $DyB_4$ is controlled by three key competing components. The first is the geometrical frustration of the Sh-S lattice with all magnetic interactions of antiferromagnetic nature. The second is the additional quadrupole degree of freedom associated with the Dy ions having a strong spin-orbit coupling. The last is the long-range RKKY magnetic exchange interaction as opposed to the nearest neighbor interaction often found in oxide materials, including SrCu2(BO3)2. As far as we are aware of, it is a theoretically untested ground as to how a particular ground state of the Sh-S lattice is selected when there are also two additional ingredients such as quadrupole order and long-range RKKY interaction.

To summarize, $DyB_4$ offers a unique opportunity for investigating the metallic Sh-S lattice with a strong Ising spin, which is coupled to the quadrupole order parameter through a strong spin-orbit coupling. Its physics is further enriched by a long-range RKKY-type exchange interaction. Put together, they produce the unique magnetic structures of $DyB_4$ both above and below the quadrupole ordering at $T_Q$=12 K. The dominant interaction is the antiferromagnetic coupling of the RKKY-type along the shortest Dy-Dy *dimer* (yellow in Fig. 3), which determines the antiparallel orientation of both the c- and ab- components of the magnetic moments of these dimmers. With the detailed structure studies, we provide a complete experimental picture of how the ground state of $DyB_4$ is selected out of otherwise degenerate competing phases with the intrinsic frustration of the Sh-S lattice. Thus these unique properties of $DyB_4$ reflect an interesting interplay between the intrinsic frustrations of the Sh-S lattice and the quadrupole ordering under the itinerant RKKY interaction.

**Acknowledgement**
We acknowledge K. B. Lee, A. Pirogov, Y. Kuramoto, Y. B. Kim, and J. van Duijn for helpful discussions and C. Y. Park for technical assistance with the XRD measurements. We would like to thank especially M. D. Le for critical reading of the manuscript and helpful comments. The work at the IBS CCES was supported by the research program of Institute for Basic Science (IBS-R009-G1) and one of us (SL) by the National Research Foundation, Korea (Grant no. NRF-2012M2A2A6004261). The work of D.Kh. was supported by Koeln University via German Excellence Initiative and by the German project SFB 1238. We acknowledge Diamond Light Source for access to Beamline I11 on proposal EE12470.

## 5. References

# Corresponding author: jgpark10@snu.ac.kr

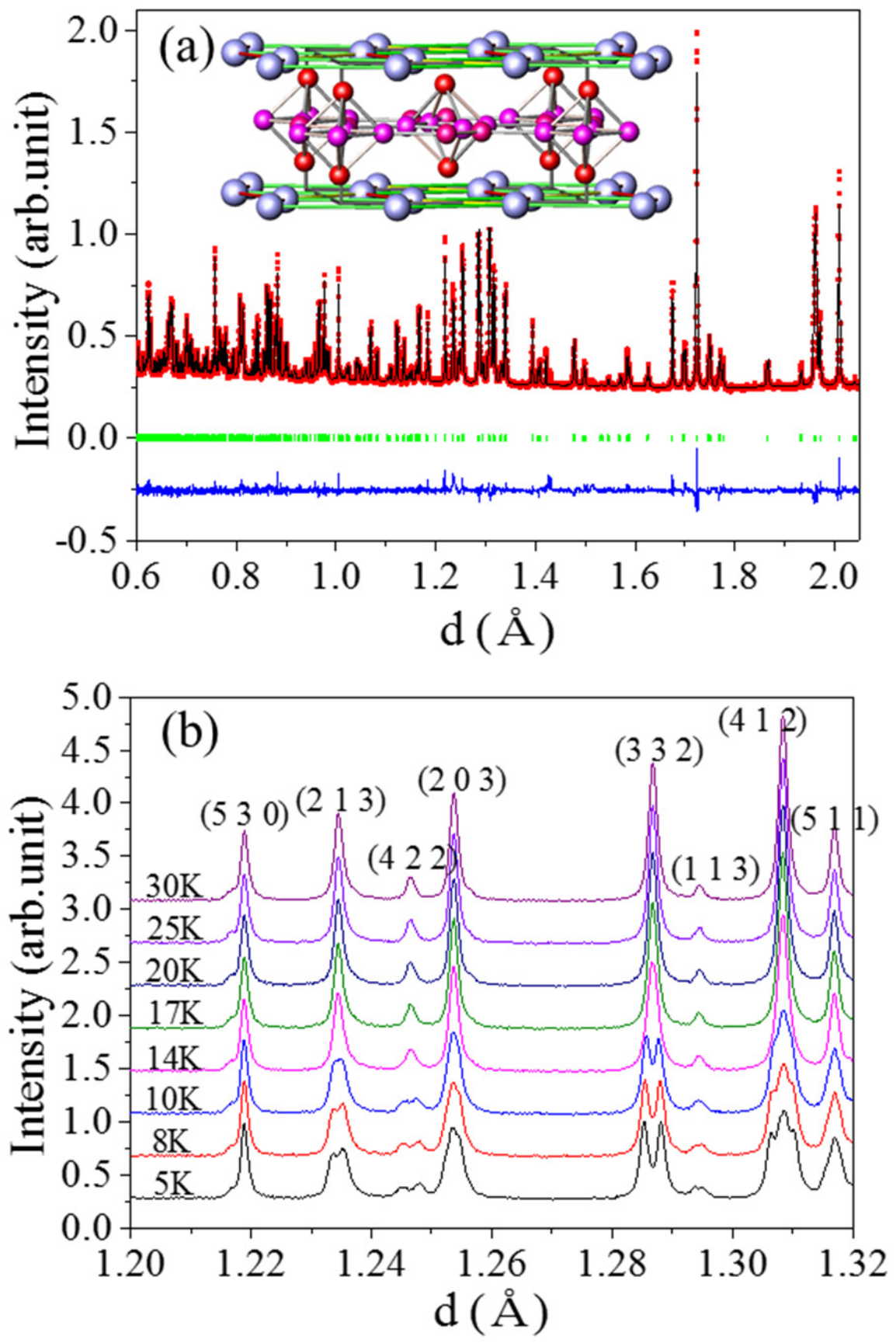

Fig. 1 (Color online) (a) The raw high-resolution neutron diffraction (symbol) data taken at 5 K together with our refinement results using a monoclinic structure (see the inset). The line underneath the data points is for our refinement results using the background shown as a black line. The vertical bars are for the positions of the Bragg peaks allowed for the monoclinic phase and the line at bottom show the difference curve. (b) The temperature evolution of the Bragg peaks from 5 to 30 K. As can be seen, several Bragg peaks: (2 1 3), (4 2 2), (3 3 2), and (4 1 2), become split below $T_Q$.

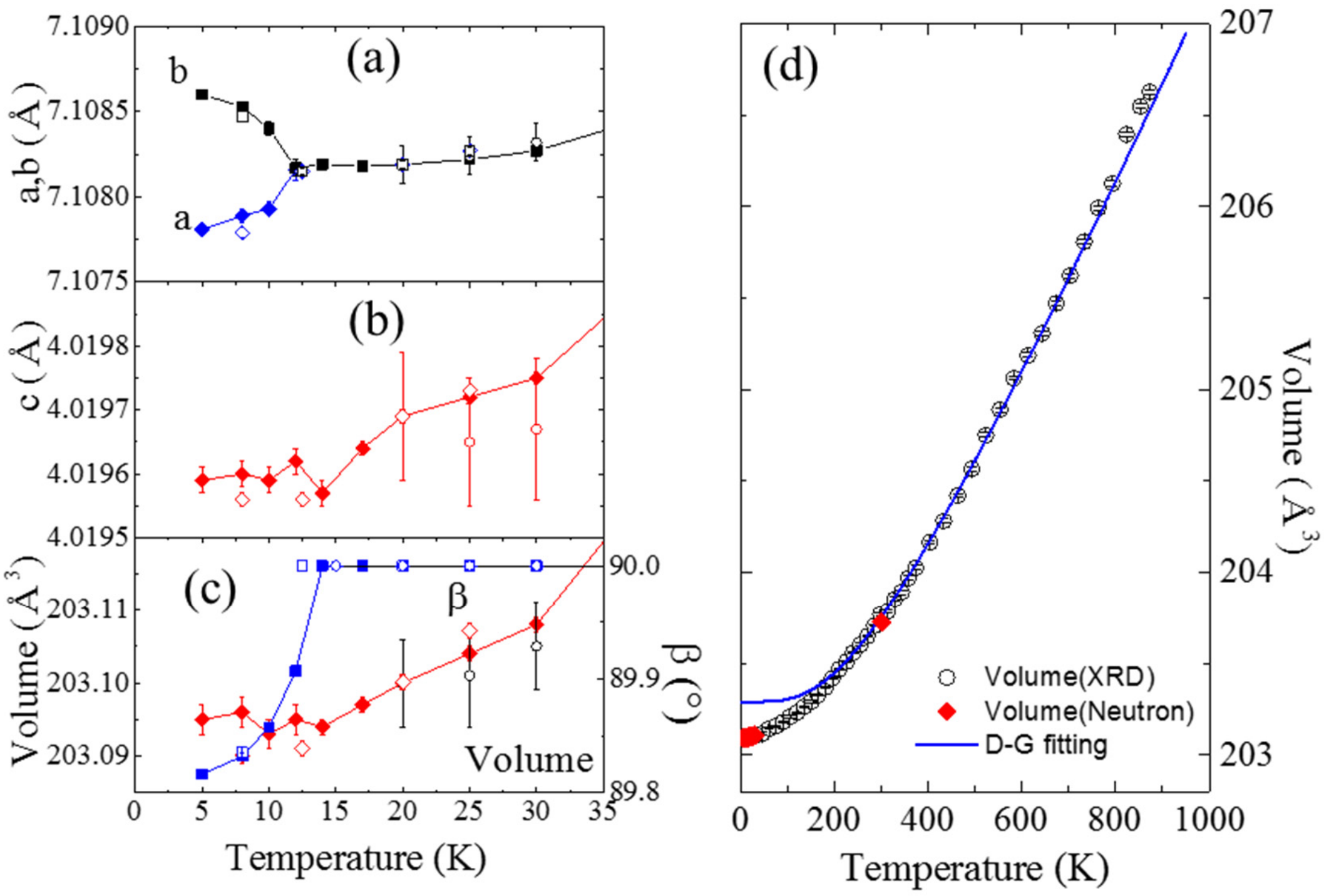

Fig. 2 (Color online) Temperature dependence of the crystal structure: (a) a and b lattice constants, (b) c lattice constant, and the unit cell volume in (c) and (d) by including data from three different experiments: neutron (filled squares & diamonds), synchrotron x-ray (open squares & diamonds), and lab-based XRD data (open circles). The line in (d) is the theoretical curve calculated by using the Debye-Grüneisen (D-G) formula as discussed in the text.

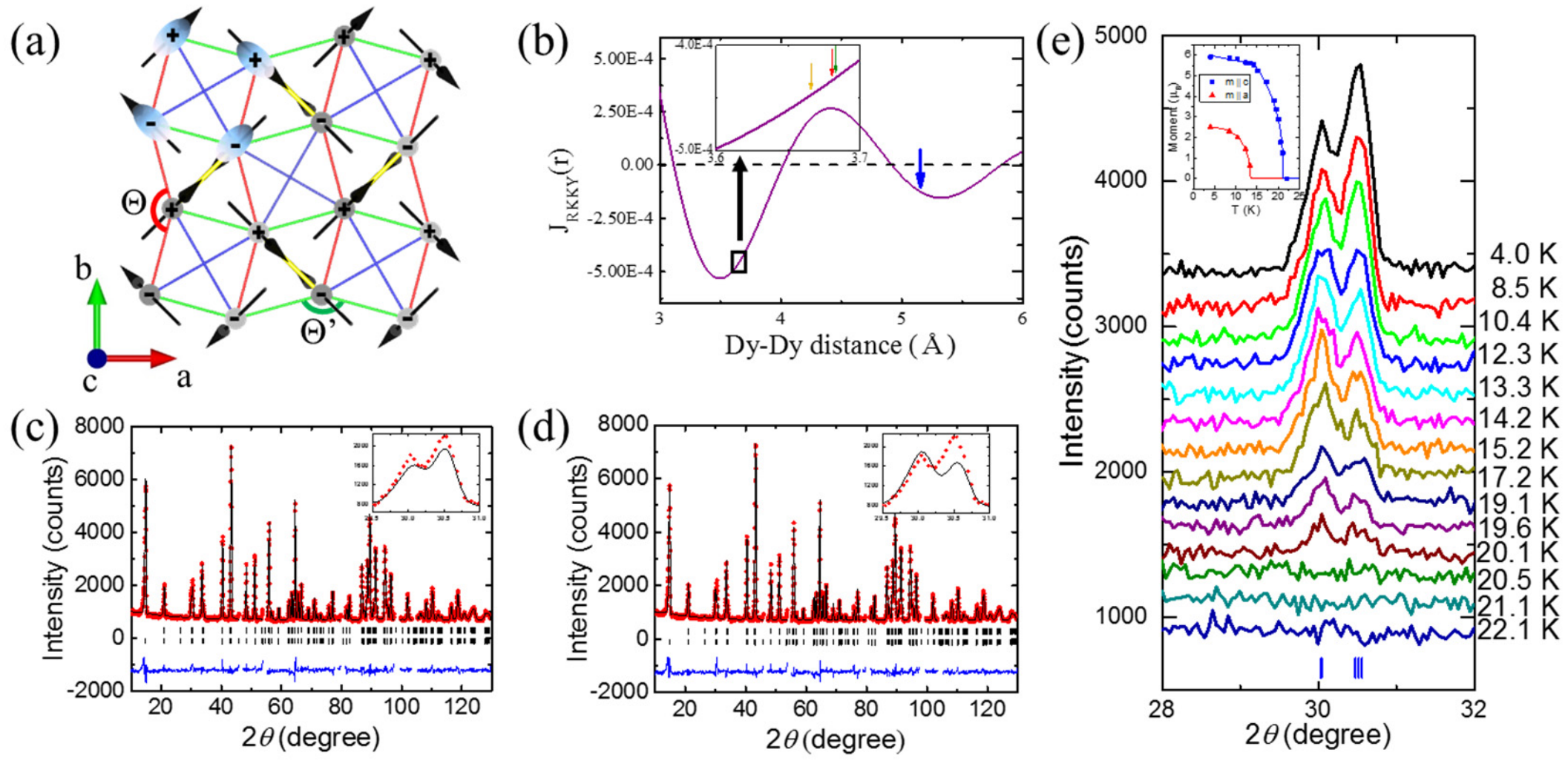

Fig. 3 (a) $\Gamma_2$ magnetic structure with four different Dy-Dy distances in the quadrupole ordered phase: the shortest yellow one (3.6498 Å), the red one (3.6617 Å), the green one (3.6716 Å), and the blue one (5.11604 and 5.2104 Å), shown in different colors. The plus (minus) signs indicate the positive (negative) c-axis magnetic component. The ellipsoids show the $Q_{zx}$-type ordering of Dy quadrupole moments. (b) Plot of RKKY exchange integral as a function of Dy-Dy distances with arrows marking the four different nearest Dy-Dy distances. Two representative results of our magnetic structure analysis at 4 K using two models: (c) $\Gamma_2$ and (d) $\Gamma_4$, as discussed in the text with the following agreement factors: $R_{mag}$= 6.55 and 9.15 % for $\Gamma_2$ and $\Gamma_4$ models, respectively while $\chi^2$ was 3.65 and 4.13 for $\Gamma_2$ and $\Gamma_4$ models, respectively. The insets in Fig. 3(c) and (d) show that the magnetic peaks can be only properly fitted by the $\Gamma_2$ model: (200) and (020) of the left peak and (101), (0-11), (011), and (-101) of the right peak. The temperature evolution of key magnetic Bragg peaks is shown in (e). Inset of Figure (e) shows the temperature dependence of the magnetic moment parallel to the a and c-axes.

Table 1 Summary of Rietveld refinement results for the S-HRPD data taken at three respective temperatures.

| 5 K (Monoclinic) | | | | | |
|---|---|---|---|---|---|
| Atom | Site | $x$ | $y$ | $z$ | B(Å$^2$) |
| Dy | 4*e* | 0.3175(2) | 0.8179(2) | -0.0004(3) | 0.156(7) |
| B1 | 4*e* | -0.0007(3) | -0.000(1) | 0.2016(4) | 0.49(1) |
| B2 | 4*e* | 0.0887(5) | 0.5868(4) | 0.5004(7) | 0.49(1) |
| B3 | 4*e* | 0.9604(5) | 0.1771(5) | 0.4996(7) | 0.49(1) |
| B4 | 4*e* | 0.3233(5) | 0.5379(4) | 0.5007(7) | 0.49(1) |
| Space group : $P$ 1 21/$a$ 1 (No. 14 $P$ $2_1/c$) | | | | | |
| $a$ = 7.10781(3) Å, $b$=7.10860(3)Å, $c$ = 4.01959(3) Å<br>$\beta$ = 89.8164°(4) Volume=203.095(2) Å$^3$ | | | | | |
| $R_p$ = 2.36 %, $R_{wp}$ = 3.22%, $\chi^2$=2.32 | | | | | |
| 17 K (Tetragonal) | | | | | |
| Atom | Site | $x$ | $y$ | $z$ | B(Å$^2$) |
| Dy | 4*g* | 0.31762(6) | 0.81762(6) | 0.00000 | 0.154(6) |
| B1 | 4*e* | 0.00000 | 0.00000 | 0.2013(4) | 0.50(1) |
| B2 | 4*h* | 0.0876(2) | 0.5876(2) | 0.50000 | 0.50(1) |
| B3 | 8*j* | 0.1769(2) | 0.0386(1) | 0.50000 | 0.50(1) |
| Space group : $P$ 4/$m$ $b$ $m$ | | | | | |
| $a$ = 7.10818(2) Å, $c$ = 4.01964(2) Å<br>Volume = 203.097(1) Å$^3$ | | | | | |
| $R_p$ = 2.64 %, $R_{wp}$ = 3.61 %, $\chi^2$=2.92 | | | | | |
| 25 K (Tetragonal) | | | | | |
| Atom | Site | $x$ | $y$ | $z$ | B(Å$^2$) |
| Dy | 4*g* | 0.31763(5) | 0.81763(5) | 0.00000 | 0.154(6) |
| B1 | 4*e* | 0.00000 | 0.00000 | 0.2019(3) | 0.50(1) |
| B2 | 4*h* | 0.0879(1) | 0.5879(1) | 0.50000 | 0.50(1) |
| B3 | 8*j* | 0.1769(1) | 0.0387(1) | 0.50000 | 0.50(1) |
| Space group : $P$ 4/$m$ $b$ $m$ | | | | | |
| $a$ = 7.10822(1) Å, $c$ = 4.01972(1) Å<br>Volume = 203.104(1) Å$^3$ | | | | | |
| $R_p$ = 2.37 %, $R_{wp}$ = 3.15 %, $\chi^2$=2.09 | | | | | |